\documentclass[fleqn,usenatbib]{mnras}

\usepackage{newtxtext,newtxmath}
\usepackage[T1]{fontenc}

\DeclareRobustCommand{\VAN}[3]{#2}
\let\VANthebibliography\thebibliography
\def\thebibliography{\DeclareRobustCommand{\VAN}[3]{##3}\VANthebibliography}

\usepackage{graphicx}	% Including figure files
\usepackage{amsmath}	% Advanced maths commands
\usepackage{hyperref}
\usepackage{diagbox}
\usepackage{xcolor}

\title[The infall region as a cosmological probe]{The infall region as a complementary probe to cluster abundance}

\author[C. T. Mpetha et al.]{
C. T. Mpetha,$^{1,2,3}$\thanks{E-mail: cmpetha@uwaterloo.ca (CTM)}
J. E. Taylor,$^{2,3}$
Y. Amoura,$^{2,3}$
R. Haggar$^{2,3}$
\\
$^{1}$Institute for Astronomy, School of Physics and Astronomy, University of Edinburgh,
Royal Observatory, Blackford Hill, Edinburgh, EH9 3HJ, United Kingdom\\
$^{2}$Waterloo Centre for Astrophysics, University of Waterloo, Waterloo, Ontario N2L 3G1, Canada \\
$^{3}$ Department of Physics and Astronomy, University of Waterloo, 200 University Avenue West, Waterloo, Ontario N2L 3G1, Canada\\
}

\date{Accepted XXX. Received YYY; in original form ZZZ}

\pubyear{2024}

\begin{document}
\label{firstpage}
\pagerange{\pageref{firstpage}--\pageref{lastpage}}
\maketitle

\begin{abstract}
Galaxy cluster abundance measurements provide a classic test of cosmology. They are most sensitive to the evolved amplitude of fluctuations, usually expressed as $S_8 = \sigma_8\sqrt{\Omega_m/0.3}$. Thus,  abundance constraints exhibit a strong degeneracy between $\sigma_8$ and $\Omega_{\rm m}$, as do other similar low-redshift tests such as cosmic shear. The mass distribution in the infall region around galaxy clusters, where material is being accreted from the surrounding field, also exhibits a cosmological dependence, but in this case it is nearly orthogonal to the $S_8$ direction in the $\Omega_m$--$\sigma_8$ plane, making it highly complementary to halo abundance or cosmic shear studies. We explore how weak lensing measurements of the infall region might be used to complement abundance studies, considering three different tests. The splashback radius is a prominent feature of the infall region; we show that detection of this feature in lensing data from the Euclid survey could independently constrain $\Omega_{\rm m}$ and $\sigma_8$ to $\pm 0.05$. Another feature, the depletion radius where the bias reaches a minimum, also shows cosmological dependence, though it is challenging to observe in practice. The strongest constraints come from direct measurements of the shear profile in the infall region at $2$--$4\,r_{200{\rm c}}$. Combining the latter with abundance constraints such as those reported from SRG$/$eROSITA should reduce the area of the error contours by an estimated factor of $1.2$ using a sample of clusters observed by the UNIONS survey, or a factor of $3$ using clusters observed by the Euclid Wide survey over a broader range of redshift.

\end{abstract}

\begin{keywords}
gravitational lensing: weak -- methods: observational -- galaxies: clusters: general -- galaxies: groups: general -- galaxies: haloes -- cosmological parameters
\end{keywords}

%%%%%%%%%%%%%%%%%%%%%%%%%%%%%%%%%%%%%%%%%%%%%%%%%%

%%%%%%%%%%%%%%%%% BODY OF PAPER %%%%%%%%%%%%%%%%%%

\section{Introduction}

Galaxy clusters provide an excellent test-bed for astrophysics and cosmology. Their overall abundance depends sensitively on the evolved amplitude of fluctuations, often expressed as $S_8 = \sigma_8\sqrt{\Omega_m/0.3}$. Because $S_8$ combines the effects of early power, characterised by the linearly extrapolated amplitude of fluctuations on 8\,$h^{-1}$\,Mpc scales, $\sigma_8$, and late-time growth, characterised by the matter density parameter $\Omega_m$, abundance measurements have a strong degeneracy in the $\Omega_m$--$\sigma_8$ plane \citep{Viana1996,abundance}, 
as do other, similar low-redshift probes of cosmological structure, such as cosmic shear \citep{Kilbinger15}. The assembly rate of clusters and their underlying dark matter halos has a different dependence on $\Omega_{\rm m}$ and $\sigma_8$ \citep{Yuba}, and could provide an alternative cosmological test if it could be quantified through observations of cluster structure, an idea first proposed in the early days of cluster cosmology \citep{structure_test1,structure_test2,structure_test3}.

A number of previous studies have investigated the link between cluster structure, assembly history, and cosmology. The density profile of a dark matter halo within its virialised region has been investigated as a cosmological probe, via measurements of concentration \citep{c_cosmo_1} or halo sparsity \citep{sparsity_orig}. Predictions in the inner parts of observed clusters remain uncertain, however, because of poorly constrained baryonic effects \cite[e.g.][]{Debackere21}. Furthermore, the constrains on $\Omega_{\rm m}$ and $\sigma_8$ from these methods have a similar degeneracy to abundance measurements \citep{c_cosmo,sparsity}, and so combining 
 either of them with abundance constraints may not lead to significant improvement. In contrast, the \textit{infall region} outside the virial radius, where matter is currently being accreted onto the halo, can reveal information about its recent growth history \citep{sp_MAR}. The current growth rate has in turn been shown to be a useful cosmological probe \citep{Yuba2}, with a degeneracy direction almost orthogonal to $S_8$ over a range of mass and redshift. Its independent measurement, in the same data used for abundance and cosmic shear studies, could break the $\Omega_{\rm m}$--$\sigma_8$ degeneracy, significantly improving the precision of abundance constraints.

The infall region includes a number of distinct features. The splashback radius $r_{\rm sp}$, the apocenter of the first orbit of material after it has been accreted into a dark matter halo \citep{Diemer14,Adhikari14,More15}, is the most widely studied feature and is sometimes proposed as a more physical boundary than the virial radius, as it mitigates the impact of pseudo-evolution \citep{pseudo,Diemer0_mass}. The so-called `depletion zone' \citep{FH} is the region around a halo from which material has been accreted, and hence is depleted with respect to the expected background.  The `depletion radius', the radius where the bias (defined below) reaches a minimum, is another possible definition  for the halo boundary \citep{halo_model_depletion}. Another feature of the depletion zone, the `inner depletion radius', has been shown to match up well with the ``optimal halo exclusion radius" defined in \cite{Garcia}. Both the splashback radius and depletion radius are sensitive to the mass accretion history of a halo \citep{sp_dependence,depletion_evolution}. There is then an interplay between the formation time, the recent  accretion history, and these features of the density profile, from which we may be able to derive cosmological information.

The splashback feature has been detected  significantly using galaxy number density profiles 
\citep[e.g.][]{sp_obs_6,sp_obs_7,sp_obs_10,sp_obs_12,sp_obs_13,sp_obs_14,eROSITA_sp}
% \citep{sp_obs_1,sp_obs_2,sp_obs_4,sp_obs_5,sp_obs_6,sp_obs_7,sp_obs_9,sp_obs_10,sp_obs_12,sp_obs_13,sp_obs_14,eROSITA_sp}
and weak lensing \citep[e.g.][]{sp_obs_6,sp_obs_8,sp_obs_13,sp_obs_15}, which has also been used to observe the depletion zone \citep{FH2}. Measuring the splashback radius through weak lensing is advantageous as it does not require knowledge of the galaxy bias, it allows the stacking of many dark matter halos, and provides mass-calibration for the observed sample. As this feature can be measured with good precision, using it as a cosmological test seems feasible. \citep{Diemer2_cosmo} made an initial study of the variation of the splashback radius with cosmology, testing {\sl WMAP} and Planck cosmologies, and models with scale-free power spectra. \cite{Roan}, henceforth \hyperlink{cite.Roan}{H24}, recently made a more detailed study of the cosmological dependence, showing that on cluster scales, the variation of $r_{\rm sp}$ is close to orthogonal to the $S_8$ direction in the $\Omega_{\rm m}$--$\sigma_8$ plane. There also has been work exploring how measurements of the splashback radius can be used to constrain alternative gravity models  \citep{sp_gravity_theory_2,sp_gravity_theory}. To our knowledge there has been no systematic investigation of how the depletion radius, or the general form of the density profile in the infall zone, depends on cosmology.

In this work, we use dark-matter-only simulations to determine how features in the infall region of cluster-mass halos vary with cosmology. We then investigate how accurately these features could be measured in present and forthcoming weak lensing surveys. Our goal is to determine the best feature(s) to measure, and also the best cluster mass and redshift range to use for cosmological tests. 
To illustrate the complementarity with abundance studies, we will focus on the recent constraints derived from clusters detected in the SRG$/$eROSITA All-Sky Survey \citep{eROSITA_abundance}. Beyond these results, there are exciting prospects for further improvements in abundance measurements from several present and forthcoming surveys, including the UNIONS \citep{UNIONS2020}, Euclid-Wide \citep{Euclid_Clusters}, and surveys by the Nancy Grace Roman Space Telescope \citep{WFIRST} and the Vera C. Rubin Observatory \citep{LSST}. 

The outline of the paper is as follows.
In Section\;\ref{sec:sims}, we describe the cosmological simulations used in this work. Section\;\ref{sec:3D} discusses the mean density profile of the infall region, its main features, and their cosmological dependence. We also describe our profile fitting process in detail. Section\;\ref{sec:2D} explains how the 3D profiles are projected to produce the 2D mass profiles measured by lensing, and how these projected profiles are then fitted. Section\;\ref{sec:results} presents forecasts for the constraining power of the infall region, and in Section\;\ref{sec:challenges} we consider some anticipated challenges in applying this test to observations. We conclude in Section\;\ref{sec:conclusions}.

\section{Simulations}
\label{sec:sims}

We use a suite of 21 dark-matter-only simulations described in  \cite{Yuba_thesis} (see also Amoura et al.~2024, in preparation). The simulated cosmologies range over the combinations of $\Omega_{\rm m}$ and $\sigma_8$ shown in Table\;\ref{tab:sims}. Other cosmological parameters are fixed, with a Hubble parameter $H_0 = 100\,h = 70\,$km$\,$s$^{-1}\,$Mpc$^{-1}$, a baryon density $\Omega_{\rm b}=0.0482$, and a spectral tilt $n_s=0.965$. 

Each simulation was performed using \textsc{Gadget 4}  \citep{Gadget4} with $1024^3$ particles in a $500\,$Mpc$/h$ box, giving a particle mass $\Omega_{\rm m} \times (3.23\times10^{10} M_{\odot})$, and a softening length of $2.5\,$kpc.  Dark matter halos were identified using the \textsc{Amiga Halo Finder} (AHF) \citep{AHF}, down to a lower mass limit of $\sim10^{12} M_{\odot}\,h^{-1}$, and their evolution was traced from  $z\!\sim\!30$ to $z=0$ in $119$ snapshots (except for two initial runs  denoted by an $o$ in Table\;\ref{tab:sims}, which only had output saved at $z=0$). 

In what follows, comoving distances are  written $c\,$Mpc$\,h^{-1}$. Halo masses $M_{200{\rm c}}$ and radii $r_{200{\rm c}}$ are defined as the mass enclosed within, and radius of, a spherical region 200 times the critical density $\rho_{c}$, respectively. 

As discussed in \hyperlink{cite.Roan}{H24}, for cluster-mass halos, baryonic effects are more or less negligible in the infall region, relative to the cosmological signal we consider. This conclusion is also supported by other works comparing dark-matter-only simulations with hydrodynamical simulations including baryons \citep[e.g.][]{Haggar_2021,Splashback_sims,FLAMINGO}. Thus, we have ignored baryonic effects in our analysis. These would be more significant on smaller mass scales, as discussed below, limiting the accuracy of the method on these scales.

\begin{table}
    \centering
      \caption{Combinations of $\Omega_{\rm m}$ and $\sigma_8$ assumed in the cosmological simulations of \protect\cite{Yuba_thesis} }%
    \begin{tabular}{|c|c|c|c|c|c|} \hline 
         \diagbox{$\sigma_8$}{$\Omega_{\rm m}$} &  0.2&  0.25&  0.3&  0.35& 0.4\\ \hline 
         1.0&  x&  &  &  & \\ \hline 
         0.95& & & & &\\\hline 
         0.9&  x&  x&  x&  x& x\\ \hline 
         0.85&  &  x&  x&  x& \\ \hline 
         0.8&  $o$&  x&  x&  x& x\\ \hline 
         0.75&  &  x&  x&  x& \\ \hline 
 0.7& x& & x& $o$&x\\ \hline
    \end{tabular}

    \label{tab:sims}%
\end{table}

\section{Features of the Infall Region}
\label{sec:3D}

\subsection{Calculating the mean density profile}
\label{sec:stacking}

Given a halo catalogue for each simulation and output redshift, we determine the mean density profile in a given mass range as follows.
The density profile of each individual halo is found by counting the number of particles in 400 logarithmically spaced shells from $r=0.01\,$Mpc$/h$ to $r=20\,$Mpc$\,h^{-1}$, and dividing their total mass by the volume of the shell. Shells are centred on the most bound particle in the halo; the potential impact of mis-centering is discussed in Section \ref{sec:challenges}. To overcome the shot noise present in individual profiles and match the quantities usually measured in lensing observations, we average them, calculating the mean density profile for all halos in a given mass range. Only halos with more than 200 particles are included, and any halo flagged as a subhalo by \textsc{AHF} is removed from the sample. Errors on the mean profile are computed using 500 bootstrap realisations. 

\subsection{Fitting the density profile}
\label{sec:fitting}

To fit the density profile in the infall region, we adopt the model of \cite{Diemer_fit}. This model is designed to fit to both the inner and outer regions of a dark matter halo in a way that reflects the dynamics of the material. It includes orbiting and infalling density terms:
\begin{align} 
        \rho(r) &= \rho_{\rm orbit}(r) + \rho_{\rm infall}(r) \\
        \rho_{\rm orbit}(r) &= \rho_s \, \exp{\left(-\frac{2}{\alpha}\left[ \left(\frac{r}{r_s}\right)^{\alpha}-1\right]-\frac{1}{\beta}\left[\left(\frac{r}{r_t}\right)^{\beta}- \left(\frac{r_s}{r_t}\right)^{\beta}\right]\right)}  \nonumber\\
         \rho_{\rm infall}(r) &= A\,\left(1+ \delta_1/\sqrt{\left(\delta_1/\delta_{\rm max}\right)^2+\left(r/r_{\rm pivot}\right)^{2s}}\right)
         \label{eq:rho_model}
    \end{align} 
The model has $9$ free parameters $\rho_s,\alpha,\beta,r_s,r_t,A, \delta_1$, and $\delta_{\rm max}$, with $r_{\rm pivot}$ being a fixed pivot scale that is set to 
\begin{equation}
r_{\rm pivot}(z) = \left(\frac{3 M}{4\pi \times 200\,\rho_{\rm m,P}(z)}\right)^{1/3} \,
\end{equation}
where $\rho_{\rm m,P}(z) = (1+z)^3\rho_{\rm m,P}(z=0) = \Omega_{\rm m}^{\rm Planck}\rho_{c,0} (1+z)^{3}$ for a Planck 2018 cosmology \citep{Planck}, and $\Omega_{\rm m}^{\rm Planck}=0.316$. When fitting stacked halos with a range of masses, $M$ is taken to be the minimum mass of this range. When fitting comoving density profiles, the $(1+z)^3$ term is dropped. In this model, the inner part of the profile is described by a modified Einasto profile \citep{Einasto} with an additional exponentially decaying truncation term, controlled by the truncation radius $r_t$. For the outer profile, $\delta_1$ and $\delta_{\rm max}$ are the overdensity at $r_{\rm pivot}$ and in the centre respectively, while $s$ is the power-law slope of the infalling term. Note that the matter density $\rho_{\rm m}$ in the original expression (Eq.\;(15) in \citealt{Diemer_fit}) has been replaced with an amplitude parameter $A$, as the value of $\rho_{\rm m}$ is not known a priori. 

Fits are performed over the range  $r=0.06-20\,$Mpc$\,h^{-1}$ when fitting the intrinsic 3D profiles, and $r=0.1-10\, r_{\rm pivot}\,$Mpc$\,h^{-1}$ when fitting mock observed 2D profiles, to avoid the impact of mis-centering (see Section \ref{sec:miscentre}). The python package \textsc{lmfit} \citep{lmfit} is used as a convenience wrapper incorporating the least-squares fitting function in \textsc{scipy} \citep{scipy} with the Trust Region Reflective method. The prior ranges for the fitted values are given in Table 1 of \cite{Diemer_fit} and we follow the fitting procedure described in Appendix A1 of that work, which is designed to avoid becoming trapped in a local minima of $\chi^2$.

\begin{figure*}
    \centering
    \includegraphics[width=\linewidth]{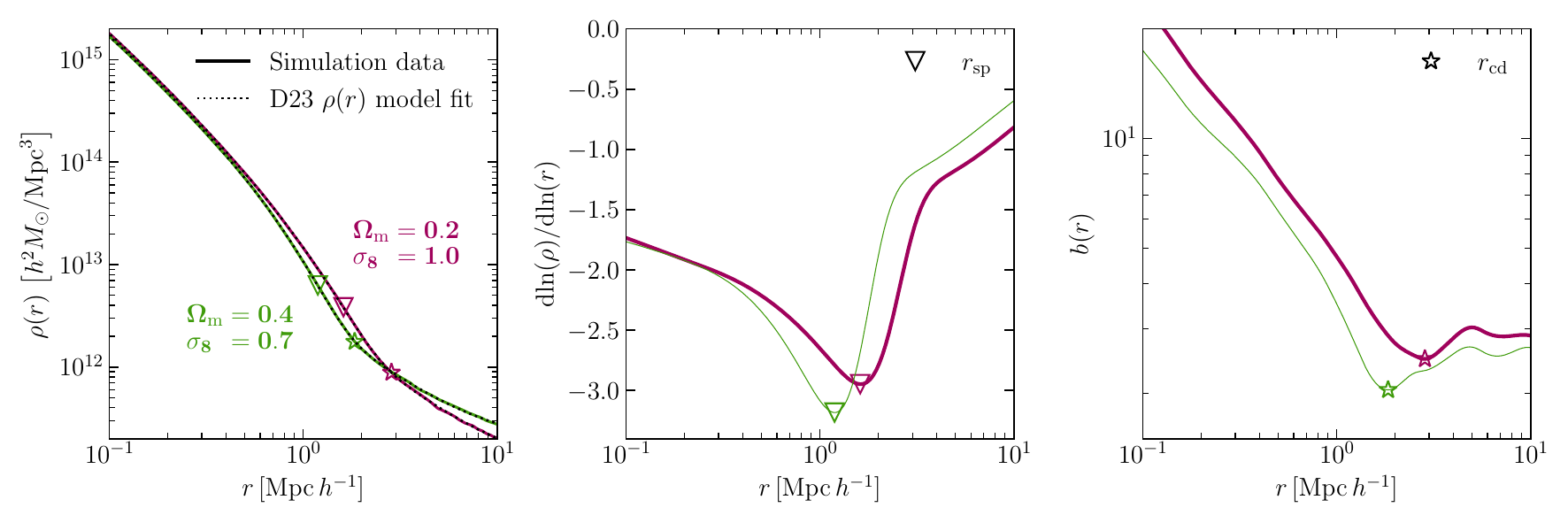}
    \caption{\textit{Left:} Mean 3D density profiles for halos with $M \geq 10^{14} M_{\odot}\,h^{-1}$ at $z=0$ (solid lines) for two simulations, with $\Omega_{\rm m}=0.2$, $\sigma_8=1.0$ (thick magenta line) and $\Omega_{\rm m}=0.4$, $\sigma_8=0.7$ (thin green line). The dotted lines show the best fit models of \protect\cite{Diemer_fit}, which are almost indistinguishable from the data. Also labelled are the splashback (triangle) and depletion (star) radii for each profile. \textit{Middle}: The logarithmic gradient $d\ln(\rho)/d\ln(r)$, whose minimum defines the splashback radius $r_{\rm sp}$. \textit{Right}: The bias profile $b(r)$, whose minimum defines the depletion radius $r_{\rm cd}$.}
    \label{fig:3Dprofiles}
\end{figure*}

Examples of mean 3D density profiles for $z=0$ halos, and the corresponding fit, can be seen in the left panel of Fig.\;\ref{fig:3Dprofiles}. 

%Figure 1

\subsection{Location of the splashback radius}
\label{sec:splashback}

Given our fits to the mean density profiles, we can proceed to test the cosmological dependence of the splashback radius and other features.
The splashback radius is identified by finding the minimum of the gradient of the logarithmic density profile, outlined in the middle panel of Fig.\;\ref{fig:3Dprofiles}. It occurs outside the virial radius $r_{200{\rm c}}$, and is dependent on the halo accretion rate \citep{sp_dependence}, which in turn is sensitive to cosmology \citep{Yuba2}. \hyperlink{cite.Roan}{H24} demonstrate that for 
cluster-mass halos, the splashback radius varies in a direction nearly orthogonal to $S_8$ in the $\Omega_{\rm m}$--$\sigma_8$ plane. The relative simplicity of identifying $r_{\rm sp}$ makes it a promising avenue to test cosmology in the infall region. The splashback feature is also easiest to observe in clusters, where the drop in density is sharper compared to lower mass halos. 

\subsection{Location of the depletion radius}
\label{sec:depletion}

The depletion radius is a more complicated feature of the infall region. The bias profile for a halo is defined by
\begin{equation}
    b(r) = \frac{\xi_{\rm hm}(r)}{\xi_{\rm mm}(r)}  = \frac{\langle\delta(r)\rangle}{\xi_{\rm mm}(r)} \, .
\end{equation}
and is related to its density profile though
\begin{equation}
    \rho(r) = \rho_{\rm m} \left[b(r)\,\xi_{\rm mm}(r)+1\right] \, .
    \label{eq:bias}
\end{equation}
Clearly, knowledge of the matter-matter correlation function is required to calculate the bias profile from the density profile, and this function is itself cosmology-dependent. The impact of uncertainty in the matter-matter correlation function on calculation of the bias profile is discussed further in Section \ref{sec:xi}. The characteristic depletion radius is given by the minimum of the bias profile, seen in the right-hand panel of Fig.\;\ref{fig:3Dprofiles}. For the cluster-mass halos shown in the figure, the bias minimum is more of a kink than a deep trough, as discussed in \cite{FH}. This flattening of the bias profile at high masses means the depletion radius is easier to measure in lower mass halos, contrary to the splashback radius. 

To avoid this problem, the minimum value of the bias profile itself could be used as the cosmological test, instead of the radius at which it reaches a minimum. However, the minimum bias is highly degenerate with the unknown value of the matter density and the poorly constrained shape of the matter-matter correlation function. For these reasons, we focus on the depletion radius, though a more sophisticated analysis that leaves $\rho_{\rm m}$ and $\xi_{\rm mm}$ free in the fitting could consider the minimum bias value.

We also note that another feature of the depletion zone is the inner depletion radius, $r_{\rm id}$, which is defined as the radius of maximum mass inflow rate \citep{FH}. To determine this radius observationally, peculiar velocities of galaxies would be required. Since we are considering an analysis based on weak lensing shear profiles alone, we do not consider this scale further.

\subsection{Cosmological dependence}
\label{sec:rcosmo}

\begin{figure}
    \centering
    \includegraphics[width=\linewidth]{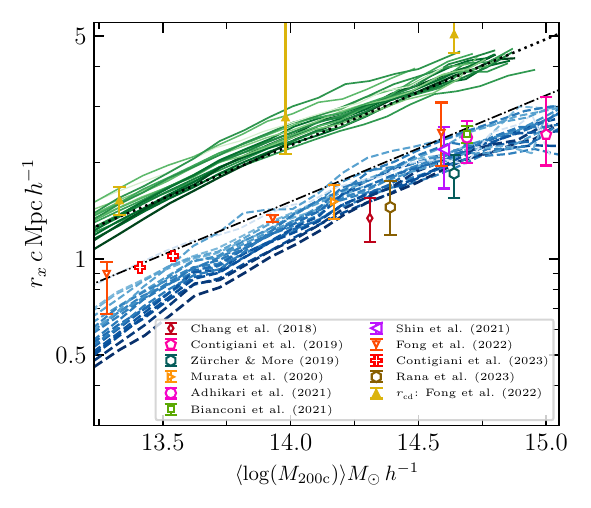}
\caption{The scaling of the splashback radius $r_{\rm sp}$ (blue dashed curves) and the depletion radius $r_{\rm cd}$ (green solid curves) with halo mass and cosmology, at $z=0$. The radii are determined by fitting the mean 3D density profile of each mass bin in the simulations. Darker line colours correspond to larger values of $S_8$, with values ranging from $0.572$ to $1.04$. Also plotted for reference are $2\, r_{200{\rm c}}$ (black dash-dot line) and $4 \, r_{200{\rm c}}$ (black dotted line). Points with error bars indicate measured values from the literature, the majority of which are of the splashback radius, with $r_{\rm cd}$ measured in one study (note these are at a range of redshifts, but have been converted to comoving units where the expected redshift variation is minimal).} 
\label{fig:r_cosmo}
\end{figure}

Fig.\;\ref{fig:r_cosmo} demonstrates how the splashback and depletion radii vary with mass and also cosmology. Halos are binned by mass, with a bin width of $0.5\,$dex, and the mean mass of the bin is plotted on the $x-$axis. The depletion radius is given by solid green lines, and the splashback radius by dashed blue lines. In both cases, darker colours correspond to larger values of $S_8$. The simulation results are shown at $z=0$, though in comoving units both radii are fairly constant with redshift, varying by less than 20\%\ between $z=0$ and $z=1$ (see Appendix \ref{app:z_var}).

Both radii vary considerably with cosmology, but the cosmological dependence also changes with halo mass. At low mass, both the splashback and depletion radii vary smoothly with $S_8$, larger $S_8$ values corresponding to smaller radii. For larger mass halos,  however, the relationship between these radii and $S_8$ is not as clear. Fig.\;$2$ of \hyperlink{cite.Roan}{H24} shows that in fact, in the cluster  mass range the splashback radius varies nearly orthogonally to $S_8$. 

%Figure 2

This mass dependence can be traced back to the halo formation time. According to hierarchical structure formation, galaxy clusters are the most recent bound structures to form in the universe, and their mean formation time varies significantly with the background cosmology.  For low$-\Omega_{\rm m}$ and high$-\sigma_8$ cosmologies, galaxy clusters typically form at earlier times \citep{Yuba}. Material currently reaching the apocentre of its orbit at the splashback radius passed through the centre of the cluster potential some time in the past, a delay we refer to as the "infall time", equal to $\frac{1}{2}$ of a radial orbital period \citep{Taylor2004}. Earlier-forming clusters will have had a larger mass in place one infall time ago, compared to later-forming clusters. A larger gravitational potential one infall time ago produces a larger splashback radius in these systems. For the depletion radius, earlier-forming clusters have had more time to draw in material from their environment, pushing the minimum of the bias profile out to larger radii. These results hint at how profile measurements in the infall region can provide constraints on cosmological parameters.

In Fig.\;\ref{fig:r_cosmo}, we also plot the examples to date of directly measured splashback radii. Masses have been converted to $M_{200{\rm c}}$ using the measured concentration when it is provided, or the concentration-mass relation of \citet{cM} when a measured value is not given, and radii have been converted to comoving units using the mean halo redshift. Current observations appear consistent with our simulations, except perhaps at the smallest mass scale. 

\section{Simulated Lensing Profiles}
\label{sec:2D}

\subsection{Projected mass density profiles}

In weak lensing measurements of mass distributions, a key observable is the reduced shear of background `source' galaxies, $g$,
\begin{equation}
    g = \frac{\gamma}{1-\kappa} \, ,
\end{equation}
where $\gamma$ is the shear and $\kappa$ is the convergence. It is often assumed in the weak lensing limit that $g \approx \gamma$ \citep{WL}, although $\gamma$ can also be inferred from $g$ through an iterative reconstruction scheme \citep{KS+}. The tangential component of the shear, $\gamma_t(\theta)$, is related to the excess surface mass density $\Delta\Sigma$ through
\begin{equation}
    \Delta\Sigma(r) = \gamma_t(\theta) \Sigma_{\rm crit} \, ,
\end{equation}
where angular separations $\theta$ on the sky are converted to radial separations using the angular diameter distance to the lens redshift $d_{A}(z)$, $r = d_{A}\theta$. The critical surface mass density is given by
\begin{equation}
        \Sigma_{\rm crit}(z_{\rm l}, z_{\rm s}) = \frac{c^2}{4\pi G}\frac{\chi_{\rm s}}{\chi_{\rm l} \left(\chi_{\rm s}-\chi_{\rm l}\right) \left(1+z_{\rm l}\right)} \, .
    \end{equation} 
The comoving distance is $\chi(z)$, where the subscript l corresponds to a foreground lens (a galaxy or a galaxy group$/$cluster), and the subscript s to a source galaxy whose shape is being measured. $\Sigma_{\rm crit}$ can be found using knowledge of the source catalogue's redshift distribution. Note that both $d_{A}$ and $\Sigma_{\rm crit}$ depend on cosmology. In our analysis, we will assume a fiducial cosmology in calculating these quantities, neglecting the cosmological variation. The impact of this approximation on cosmological constraints will be discussed in more detail, and shown to be minimal, in Section \ref{sec:cosmo_assume}.

%Figure 4
\begin{figure*}
    \centering
    \includegraphics[width=\linewidth]{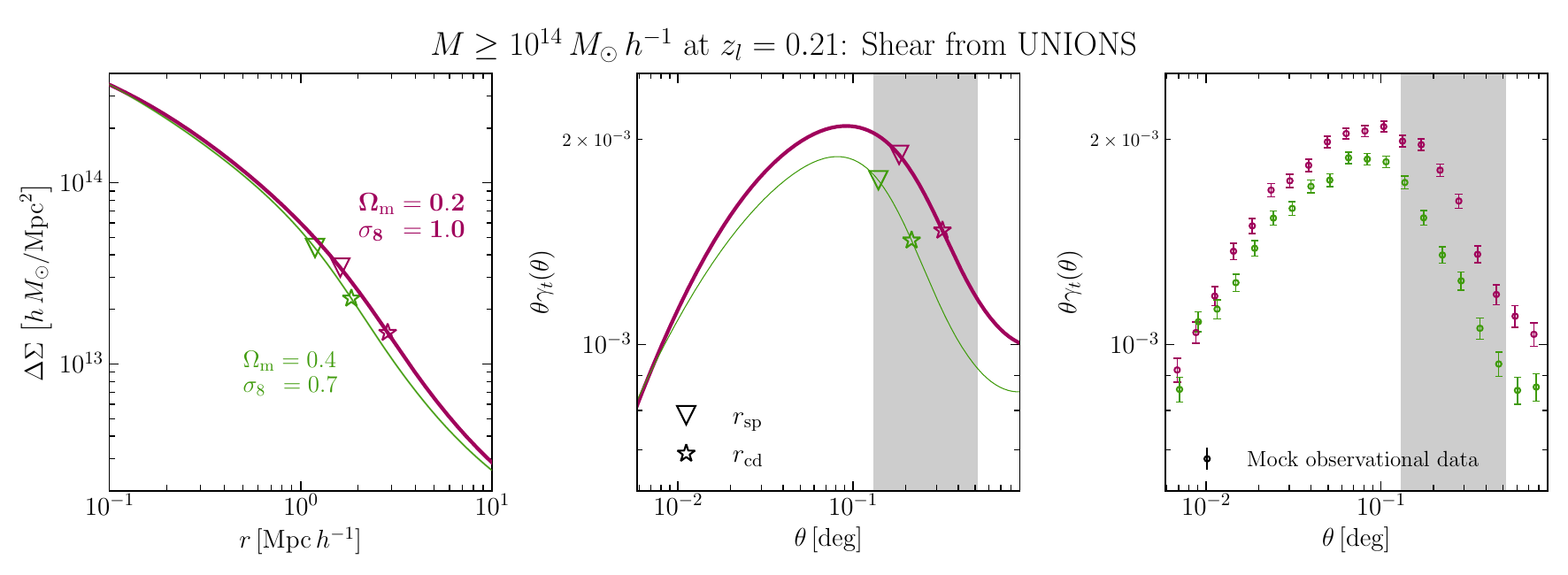}
    \caption{\textit{Left:} 3D density profiles are converted into $\Delta\Sigma$ profiles, the excess surface mass density. \textit{Middle:} The tangential shear $\gamma_{t} = \Delta\Sigma / \Sigma_{\rm crit}$ is found by assuming a lens redshift of $0.21$ and UNIONS survey source galaxy properties. Radii are converted into angles, assuming a fiducial cosmology. \textit{Right:} Mock observational data for both simulations using realistic errors and Gaussian scatter.}
    \label{fig:2Dprofiles}
\end{figure*}

Fig.\;\ref{fig:2Dprofiles} illustrates the  steps involved in going from the density profiles measured in the simulations to a predicted $\gamma$ profile. We show results for two cosmologies, one with $\Omega_{\rm m}=0.4$, $\sigma_8=0.7$ (thin green line), and another with $\Omega_{\rm m}=0.2$, $\sigma_8=1.0$ (thick magenta line). These correspond roughly to the ends of the `banana', the typical uncertainty contour derived in previous cluster abundance or cosmic shear studies. 

The stacked density profile is converted into a 2D projected surface mass density profile using an Abel transform,
 \begin{equation}
        \Sigma(R) = 2\int_R^\infty \frac{\rho(r)r}{\sqrt{r^2 - R^2}}\, dr \, ,
        \label{eq:sigma}
    \end{equation} 
where $R$ is the projected distance from the centre. Given the surface mass density distribution, the \textit{excess} surface mass density around a dark matter halo $\Delta\Sigma(R)$ is defined as
\begin{equation}
        \Delta \Sigma(r) = \overline{\Sigma}(R)-\Sigma(R) \, ,
        \label{eq:dSigma}
    \end{equation}
with
\begin{equation}
        \overline{\Sigma}(R) = \frac{2}{R^2} \int_0^R R' \Sigma(R') \, dR' \, .
    \end{equation}
This quantity is plotted in the left panel of Fig.\;\ref{fig:2Dprofiles}. Next, using an assumed lens redshift and the source redshift distribution for a given weak lensing survey (in this case UNIONS), the $\Delta\Sigma(r)$ profile is converted to a shear profile, and radial separations are converted to angular separations on the sky. This gives the $\gamma_t(\theta)$ profile in the middle panel. Finally, realistic uncertainties on the $\gamma$ profile are calculated using the \textsc{cluster-lensing-cov} package\footnote{\url{https://github.com/hywu/cluster-lensing-cov}} \citep{Wu19} and Gaussian scatter of data points is added to obtain the mock observations in the right-hand panel. We see a clear difference between the predicted shear profiles for the two cosmologies, and 
the amplitude of the difference greatly exceeds the  noise, particularly in the infall region. Thus, measurements of the infall region with a lensing survey such as UNIONS should easily distinguish between the two `ends of the banana'.

\subsection{Fitting an observed shear profile}
\label{sec:fit2D}

To pick out features in the projected, 2D density profile, we fit the observed shear profile $\gamma_t(\theta)$ using a similar approach to our 3D fitting. Starting with a model 3D profile given by Eq.(\;\ref{eq:rho_model}, we project and convert this as described above to generate the corresponding $\gamma_t(\theta)$ profile. We then compare this to the mean profile measured in the simulation using least squares, and iterate over this process to determine the best choice of 3D parameters.
Uncertainties in the fit are propagated along with simulated measurement uncertainties when generating the results in Section \ref{sec:results}. Physical units are used when stacking the 2D profiles. 20 uniformly spaced data points are assumed in the fitting range, $r=0.1-10\, r_{\rm pivot}\,$Mpc$\,h^{-1}$.

When fitting the density profiles directly to infer the 3D $r_{\rm sp}$ and $r_{\rm cd}$ from the simulated halos, $A$ in Eq.\;(\ref{eq:rho_model}) is left free, giving the most freedom to minimise the residual. 
Allowing $A$ to be free introduces strong correlation between parameters, however, so when noisy stacked lensing profiles are being fitted its value is fixed to $\rho_{\rm m}^{\rm Planck}$ to avoid large parameter uncertainties caused by strong degeneracies, at the cost of larger residuals.

\section{Cosmological Constraints from the Infall Region}
\label{sec:results}

As demonstrated in \hyperlink{cite.Roan}{H24} and Figs \;\ref{fig:3Dprofiles}--\ref{fig:2Dprofiles}, the density profile in the infall region of clusters shows clear systematic variations with cosmology. We will now investigate the prospects of using this dependence to constrain $\Omega_{\rm m}$ and $\sigma_8$, given lensing data from current and forthcoming surveys. For each specific survey considered, the uncertainty on shear profiles depends on the properties of an assumed lens sample (number of halos, mass and redshift distributions) and source sample (shape noise, surface density, and redshift distribution). Our assumed source and lens properties are outlined in Section \ref{sec:surveys}. In Sections  \ref{sec:depletion_results} and
\ref{sec:splashback_results} respectively, 
we then consider constraints obtained by measuring the depletion radius or the splashback radius. Finally, in Section \ref{sec:fullprofile} we estimate constraints based on fitting the entire infall region.

Throughout this section, covariances are calculated using the \textsc{cluster-lensing-cov} package \citep{Wu19}. Uncertainties on the profile are used to generate Gaussian scatter in the data points. To calculate our final results (either fitting the profile to extract the splashback/depletion radius, or comparing the shear profiles directly), we repeat each analysis 50 times with random realisations of the errors, and average these multiple realisations to give the final quoted value.

\subsection{Source and lens catalogue assumptions}
\label{sec:surveys}

We focus on two weak lensing surveys, UNIONS and Euclid Wide. The mass and redshift bins for the cluster sample assumed in each survey are summarised in Table\;\ref{tab:survey_bins}.

\subsubsection{UNIONS}
UNIONS \citep{UNIONS2020} is a 5-band ground-based imaging survey. The weak lensing component of the survey will cover an area of $4800\,$deg$^2$, with source number density of $10\,$deg$^{-2}$ and shape noise of $0.34$, as described in \cite{shapefit}. The source redshift distribution $n(z)$ from that work is also assumed here. For the lens catalogue, we use the sample described in \cite{Spitzer_thesis} as a reference case. This was based on halos in the UNIONS footprint taken from the catalogue created by \cite{Tinker}, where a halo finder was run on SDSS galaxies. However \cite{Spitzer_thesis} considered only $2000\,$deg$^2$ of UNIONS, as the weak lensing survey was not complete at that time. We have assumed that the mean number density of lenses is the same across the full $4800\,$deg$^2$ of the survey, and therefore lens numbers in that work are scaled up by a factor of $2.4$. Our three lens mass bins also correspond to those defined in \cite{Spitzer_thesis}.

\subsubsection{Euclid}
The Euclid Wide survey, planned for the Euclid telescope, is a space-based survey covering $36\%$ of the sky, providing a total area of $15000\,$deg$^2$, with a source number density of $30\,$deg$^{-2}$ and shape noise of $0.3$ \citep{Euclid_WL}. The mean redshift of the source distribution is $\langle z \rangle=0.95$. As lenses, we consider cluster-mass halos detected in the Euclid photometric survey. The numbers in Bins $1-5$ of Table\;\ref{tab:survey_bins} are derived from Fig.\;3 of \cite{Euclid_Clusters}, assuming a detection threshold of clusters with $N_{500{\rm c,field}}/\sigma_{\rm field}=5$. The masses are based on the corresponding selection function in Fig.\;2 of that work, converted to $M_{\odot}\,h^{-1}$ assuming $h=0.7$. 

We focus on the lower redshift bins of the survey, to avoid incompleteness which might bias cluster properties and the inferred shape of the infall region. We also consider a group-scale sample in Bin 0, assuming a complete halo sample in the mass range $10^{13} \leq M\,[M_{\odot}\,h^{-1}] \leq 10^{13.5}$ with a total number calculated from the halo mass function of \cite{despali} for a Planck cosmology. The resulting sample contains $\sim \! 10^5$ halos, and represents the best we could hope to do in this mass range.

\begin{table}
    \centering
        \caption{Mass and redshift bins assumed for the UNIONS and Euclid-wide weak lensing surveys. Columns give the bin number, mean redshift, mass range and number of clusters respectively. Each redshift shell spans $\pm0.05$ around the mean value.}
    \begin{tabular}{c|lll|lll}
         &   \multicolumn{3}{c}{UNIONS}&\multicolumn{3}{c}{Euclid}\\
 Bin& $\langle z_{\rm l}\rangle$&$M/M_{\odot}\,h^{-1}$& $N_{\rm l}$& $\langle z_{\rm l}\rangle$& $M/M_{\odot}\,h^{-1}$&$N_{\rm l}$\\  
        \hline 
 \rule{0pt}{3ex}  0& $0.15$& $10^{13-13.5}$& $19745$& $0.2$& $10^{13-13.5}$&$10^5$\\
         1& 
       $0.17$&$10^{13.5-14.0}$&$9199$& $0.25$& $\geq10^{14.3}$&$3500$\\
 2&   $0.21$&$\geq10^{14.0}$&$3919$& $0.35$& $\geq10^{14.2}$&$7000$\\
 3& \multicolumn{3}{c|}{---}& $0.45$& $\geq10^{14.1}$&$12500$\\
 4& \multicolumn{3}{c|}{---}& $0.55$& $\geq10^{14.0}$&$20000$\\
 5&   \multicolumn{3}{c|}{---}& $0.65$&$\geq10^{14.0}$&$25000$\\\end{tabular}
    \label{tab:survey_bins}
\end{table}

\subsection{Constraints from the depletion radius}
 \label{sec:depletion_results}

The depletion radius is defined as the point where the bias profile reaches a minimum value. To measure this, we fit the stacked $\Delta\Sigma(R)$ profiles with the density profile model. Errors on $\Delta\Sigma(R)$ are found using the \textsc{cluster-lensing-cov} package, as described in the previous section. These errors are propagated through the fitting procedure. The result is a set of best-fit parameters for the density profile model, and their covariance. The covariance is used to generate a set of correlated samples of the model parameters, assuming Gaussian uncertainties. Then a simple Monte Carlo operation is performed, drawing from this correlated set of samples repeatedly, and finding the new depletion radius of the density profile created on each random draw.

Once a density profile is fitted from a measured $\gamma$ profile, there are still two unknowns in Eq.\;(\ref{eq:bias}); the matter density $\rho_{\rm m}$, and the matter-matter correlation function $\xi_{\rm mm}(r)$, which is calculated using \textsc{colossus} \citep{colossus}. We assume a Planck cosmology for both of these, as in real data we will not know the true value of $\rho_{\rm m}$, and while $\xi_{\rm mm}(r)$ can be determined from the observations, it requires knowledge of the galaxy bias. The net effect of this approximation will be discussed further in Section\;\ref{sec:xi}.

For the highest mass halos in our sample, there is no actual dip in the bias profile around the depletion radius, as expected from \cite{FH}. These halos form in high-density environments, where the bias profile remains positive at all radii. To test the use of the depletion radius as a cosmological probe, we therefore consider a sample of less massive halos (`Bin 0', described above). At present, the depletion radius for halos in this mass range has been measured through weak lensing by \cite{FH2}, to a precision of $\Delta r_{\rm cd} = 0.14$. 

The top-left and bottom-left panels of Fig.\;\ref{fig:SNR_all} show predicted uncertainties on the depletion radius in a stacked sample of group-mass halos, with profiles observed using UNIONS and Euclid respectively. With Euclid data, our predicted uncertainties lie in the range $0.1$-$0.15$, making them consistent with the only existing measurement in the literature \citep{FH2}. The predicted uncertainty in this quantity is larger than the cosmological variation over the range of $\Omega_{\rm m}, \sigma_8$ considered, as shown in the left-hand columns of Fig.\;\ref{fig:SNR_all}. We conclude that the depletion radius is not a good candidate for deriving cosmological constraints. The assumptions made when finding the bias profile also lead to a mismatch between the true and inferred bias minimum in some cases. This is discussed in more detail in Section \ref{sec:xi}, and is a serious limitation to using the location of the depletion radius as a cosmological test.

\subsection{Constraints from the splashback radius}
\label{sec:splashback_results}

To determine the splashback radius, we use a similar approach as in Section \ref{sec:depletion_results}. We focus on cluster-mass halos for this measurement, as they will provide the largest SNR. We note that \hyperlink{cite.Roan}{H24} find that the splashback radius needs to be measured to an accuracy of $\sim\!5\%$ if it is to be a competitive independent probe on $\Omega_{\rm m}$ and $\sigma_8$. Other works have shown it is observable in the future to relatively high precision; for example \citep{sp_measure} find an uncertainty on $r_{\rm sp}$ of $\sim0.05$ ($3-6\%$ depending on the value) from forthcoming weak lensing surveys such as Euclid and LSST \citep{LSST}. 

The top-middle and bottom-middle panels of Fig.\;\ref{fig:SNR_all} summarise our results for profiles of stacked cluster-mass samples, measured using lensing data from UNIONS and Euclid respectively. For each simulation, the value of the splashback radius and its associated uncertainty is shown. For UNIONS, the errorbars are too large to constrain deviations from a Planck-like cosmology (indicated by the magenta band, with the width indicating the measurement error for that cosmology). The reported values from Planck are $\Omega_{\rm m}=0.316$ and $\sigma_8=0.811$ \citep{Planck}. On the other hand for Euclid, depending on the cosmology, we see that significant deviations from the expected Planck value can be observed. Overall, we find that deviations from the fiducial Planck values of  $\Omega_{\rm m}$ and $\sigma_8$ can be constrained to $\pm0.05$, using splashback radius measurements for the Euclid cluster sample. 

These results can be further improved by measuring $r_{\rm sp}$ in stacked clusters from the four other redshift bins in Euclid in Table\;\ref{tab:survey_bins}. The cosmological dependence tends to be similar or larger in higher redshift bins, and so for each simulation, stacking the SNR from all five bins improves the constraining power by at least a factor of two, even reaching a factor $4$--$6$ improvement for cosmologies in close proximity to Planck in $\Omega_{\rm m}$--$\sigma_8$. We note this does require complete samples and accurate mass estimates for clusters at redshifts $\geq 0.3$.

Our results mirror those of \cite{sp_measure}: the projection effects involved in measuring a $\Delta\Sigma$ profile with weak lensing weaken constraints on the splashback radius. The quoted uncertainties on $r_{\rm sp}$ in that work are smaller than here, likely due to the different number of halos assumed ($25\,000$ in their case, compared $3\,500$ in ours). Previous work has also demonstrated a bias on the inferred splashback radius in optically selected cluster catalogues \citep{rsp_bias1,rsp_bias2}. These selection effects would need to be modelled with more realistic simulations, depending on the cluster sample assumed.

\subsection{Constraints from the full profile}
\label{sec:fullprofile}

Fitting a model of the density profile to observations in order to identify characteristic radii introduces large uncertainties. To avoid this, we can consider comparing an observed shear profile directly to simulated profiles, without assuming any particular analytic form. To evaluate the prospects for this method, we use the signal-to-noise ratio (SNR) of the {\it difference} between the average shear profiles in the infall region for two different cosmologies. Given cosmologies $A$ and $B$, 
\begin{align}
    {\rm SNR} &= \frac{\lvert \langle \gamma_t({\rm infall}) \rangle_A - \langle \gamma_t({\rm infall}) \rangle_B \rvert}{\sqrt{\sigma_A^2 + \sigma_B^2}} \, , \\
    \sigma &= \frac{1}{N}\sqrt{\sum \sigma_i^2}\, ,
    \label{eq:SNR}
\end{align}
where $\sigma_i$ is the uncertainty on the $i^{\rm th}$ data point and $N$ is the number of data points in the annulus. The average is computed in an angular annulus encompassing the infall region, described in the next section.  Before performing this test, we first need to identify the angular scale where the SNR of the difference is greatest.

\subsubsection{Scale dependence of the SNR}
\label{sec:annulus}

\begin{figure}
\centering
    \includegraphics[width=\linewidth]{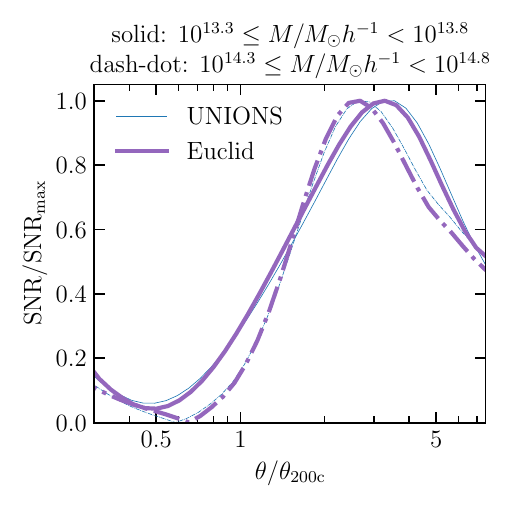}
    \caption{The SNR of the difference in stacked shear profiles for two cosmologies, as a function of angular distance from the halo centre, in units of the virial radius $\theta_{200{\rm c}}$. For this demonstrative example, the difference is between the mean shear profiles in cosmology $A$ with $\Omega_{\rm m}=0.25$, $\sigma_8=0.85$ and those from cosmology $B$ with $\Omega_{\rm m}=0.35$, $\sigma_8=0.75$. The example shown uses the UNIONS bin with $0.16\leq z_{\rm l}<0.26$ (thin blue line) and the Euclid bin with $0.2 \leq z_{\rm l} < 0.3$ (thick purple line). Solid and dashed-dot lines correspond to group-mass halos and cluster-mass halos, respectively. In each case, the SNR has been normalised by its maximum value.}
    \label{fig:SNR_R}
\end{figure}

To test how the SNR of the cosmological variation of the profile depends on distance from the halo centre, we calculate mean profiles for two cosmologies, `$A$' with $\Omega_{\rm m}=0.25$, $\sigma_8=0.85$ and `$B$' with $\Omega_{\rm m}=0.35$, $\sigma_8=0.75$, and take their difference. These two cases were chosen as they bracket the Planck values in the $\Omega_{\rm m}$--$\sigma_8$ plane along the $S_8$ degeneracy direction, the direction in which the profile of the infall region is expected to vary the most. The angular dependence of the SNR of this difference, normalised by its maximum value, is shown in Fig.\;\ref{fig:SNR_R}. The SNR depends on the source and lens redshift distributions, and is shown for two examples, the UNIONS bin with $0.16\leq z_{\rm l}<0.26$ (thin blue line) and the Euclid bin with $0.2 \leq z_{\rm l} < 0.3$ (thick purple line). Solid and dashed-dot lines correspond to group-mass halos and cluster-mass halos, respectively.

The SNR peaks in the infall region, indicating that this is the best place to look for differences caused by the cosmological parameters. The smooth peak at $\sim\!2$--$4\,\theta_{200{\rm c}}$ persists when comparing differences between other cosmologies and considering different mass bins, though it does vary slightly with mass, and typically has a width of $\sim\!0.15-0.25\,$dex. To determine the best angular scale to use for a given sample and survey, we measure the location of the peak in SNR for all combinations of cosmologies and over a large mass range, and determine the mean $\theta_{\rm peak}$--mass relation. It is given by
\begin{equation}
    {\rm log}(\theta_{\rm peak} / {\rm deg}) = a \, {\rm log}(M / M_{\odot}\,h^{-1}) + b \, ,
\end{equation}
where for UNIONS $a=0.21$, $b=-3.6$, and for Euclid $a=0.36$, $b=-5.9$. The final annulus over which we integrate has a width of $\pm0.2\,$dex around the peak position. For any mass bin, the adopted annulus always encompasses the peak of SNR, and it is not cosmology dependent. This annulus is used to produce the results in the top-right and bottom-right panels of Fig.\;\ref{fig:SNR_all}, for UNIONS and Euclid respectively.

%Figure 7

\subsubsection{Dependence of the profile variation on mass and redshift}
\label{sec:SNR-mass-z}

\begin{figure}
\centering
    \includegraphics[width=\linewidth]{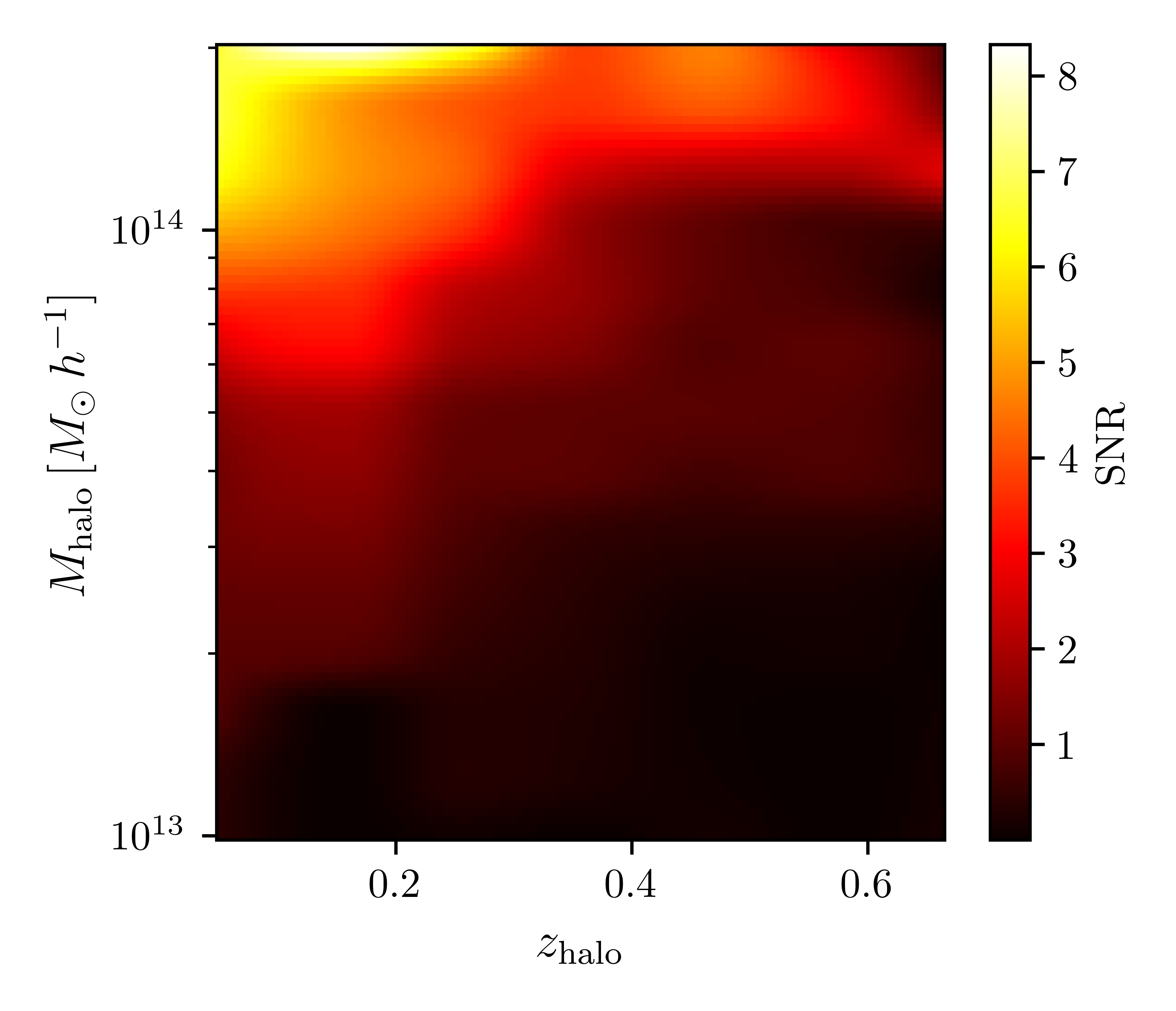}  \caption{The SNR of the difference in the predicted mean shear profile between two cosmologies, $A$ with $\Omega_{\rm m}=0.25$, $\sigma_8=0.85$ and $B$ with $\Omega_{\rm m}=0.35$, $\sigma_8=0.75$. The SNR is calculated in nine redshift bins and fourteen mass bins, with a fixed $1000$ halos per bin.}
    \label{fig:SNR_heat}
\end{figure}

Given the mass-dependent radial annulus described in the previous section, we calculate the average $\gamma_t$ within that annulus for the two different cosmologies, and determine the SNR of the difference between them. Fig.\;\ref{fig:SNR_heat} shows this SNR as a function of mass and redshift, assuming a fixed $1000$ halos per mass and redshift bin, and Euclid weak lensing data. 
%Figure 8

\begin{figure}
    \centering
    \includegraphics[width=\linewidth]{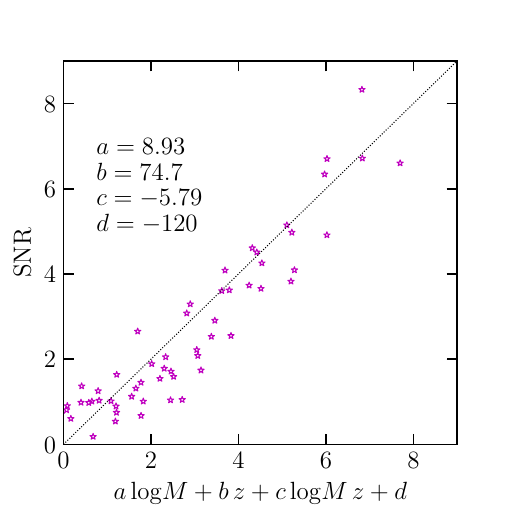}
    \caption{Bi-linear fit to the SNR in Fig.\;\ref{fig:SNR_heat} (dotted line), together with residuals with respect to this fit (stars).}
    \label{fig:SNR_fit}
\end{figure}

To get a rough sense of variation of the SNR with mass and redshift, we make a simple bi-linear fit to the values in Fig.\;\ref{fig:SNR_heat}:
\begin{align}
    f({\rm log}\left(M/M_{\odot}\,h^{-1}\right), z) =& 8.93\,{\rm log}\left(M/M_{\odot}\,h^{-1}\right) + 74.7\,z \\ & - 5.79\, {\rm log}\left(M/M_{\odot}\,h^{-1}\right)\,z -120 \, . \nonumber
\end{align}
An inverse weighting of $1/$SNR is used in the fit to prioritise the high-SNR regime. More complicated functions could better fit both high and low SNRs, but here we opt for this form for simplicity. The fit and residuals are shown in Fig.\;\ref{fig:SNR_fit}

Ignoring systematic errors and biases, and assuming the SNR scales as SNR$\,\propto \sqrt{N}$, we can use these results to set requirements on the sample size needed to distinguish cosmological models. For example, to distinguish between the two cosmologies considered so far ($A$ with $\Omega_{\rm m}=0.25$, $\sigma_8=0.85$ and $B$ with $\Omega_{\rm m}=0.35$, $\sigma_8=0.75$) at $5\sigma$ significance, assuming clusters with a mean mass of $10^{14} M_{\odot}\, h^{-1}$ at $\langle z_{\rm l} \rangle \!\sim\!0.2$, we would need a sample size of $\sim\!1500$ clusters.

\subsection{Summary}

\begin{figure*}
    \centering
    \includegraphics[width=0.85\linewidth]{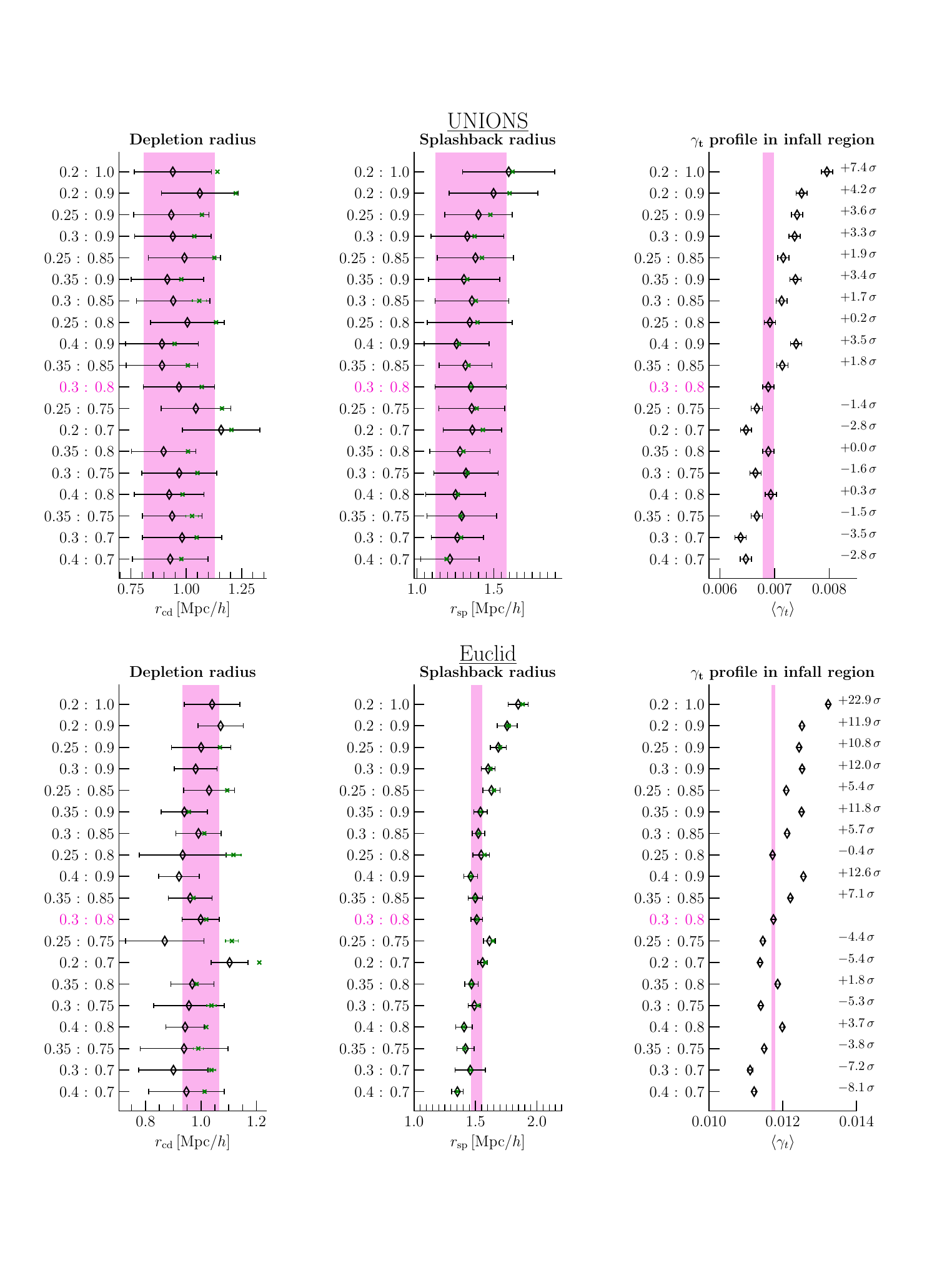}
    \caption{A summary of our results for two weak lensing surveys, UNIONS (top three panels), and Euclid (bottom three panels). From left to right, we show predictions for the measured characteristic depletion radius $r_{\rm cd}$, the splashback radius $r_{\rm sp}$, and the amplitude of the shear profile in the infall region for 19 different combinations of $\Omega_{\rm m}$ and $\sigma_8$. In each case, the value measured in the combination of parameters most similar to a Planck cosmology, $\Omega_{\rm m}=0.3$ and $\sigma_8=0.8$, is highlighted with a purple band. The true 3D $r_{\rm sp}$ and $r_{\rm cd}$ is overlaid with a green symbol for each simulation.}
    \label{fig:SNR_all}
\end{figure*}

Fig.\;\ref{fig:SNR_all} summarises the expected accuracy of cosmological constraints derived using the three methods, depletion radius measurements, splashback radius measurements, or full profile measurements, for the UNIONS survey (top panels) and the Euclid Wide survey (bottom panels). The vertical ordering of cosmologies in each panel is based on the fitting function in Fig.\;6 of \hyperlink{cite.Roan}{H24}. To derive these results, we have considered only the lenses in the highest mass Bin of Table\;\ref{tab:survey_bins} for each survey. Using additional bins would improve these constraints further.

\begin{figure*}
\centering
    \includegraphics[width=\linewidth]{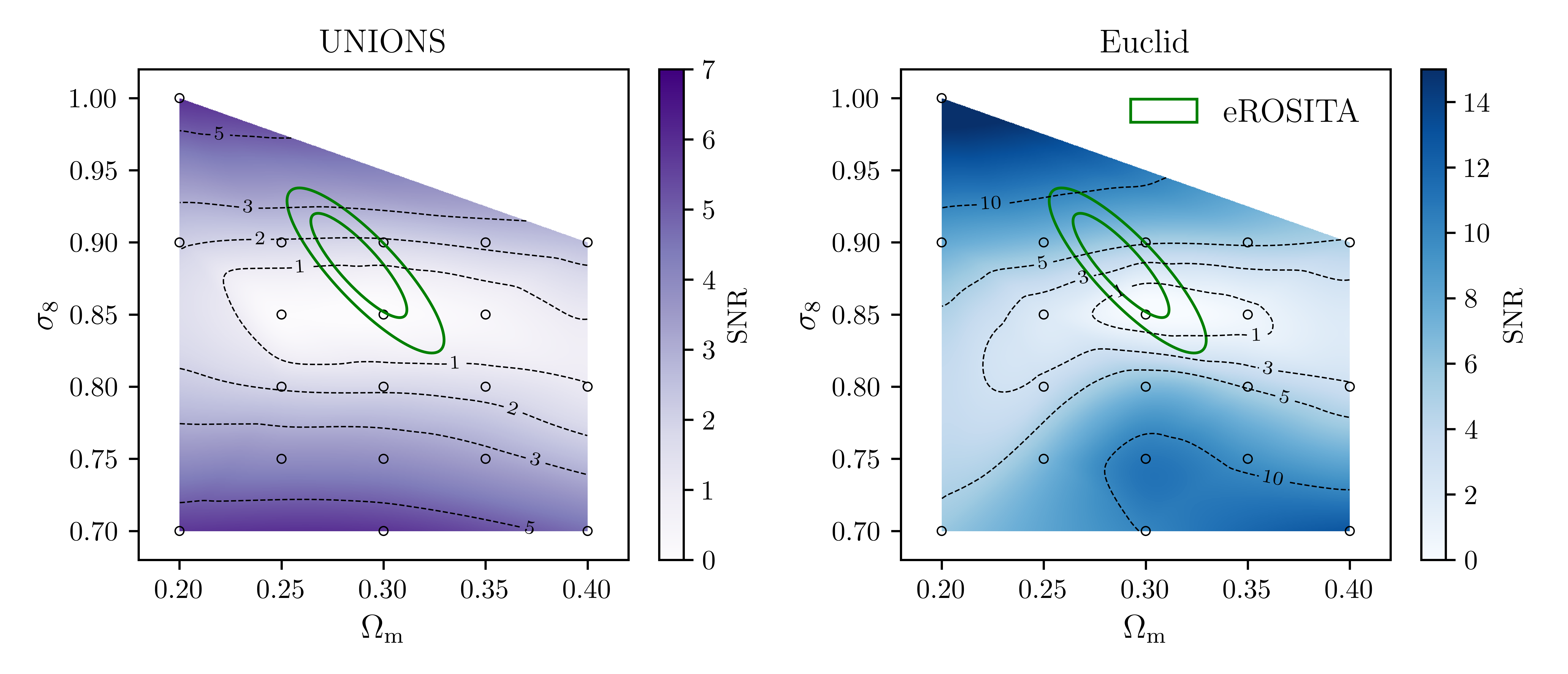}
    \caption{SNR with which deviations from a fiducial cosmology could be detected with the full profile measurement method, assuming lensing measurements from UNIONS (left-hand panel) and Euclid (right-hand panel). In each case, only the single most massive bin of clusters is used. The fiducial cosmology is taken to be $\Omega_{\rm m}=0.3$, $\sigma_8=0.85$. Approximate $1$ and $2$--$\sigma$ abundance constraints from \protect\cite{eROSITA_abundance} are shown for comparison (green ellipses).}
    \label{fig:SNR}
\end{figure*}

%Figure 9
Fig.\;\ref{fig:SNR} plots the results of the top-right and bottom-right panels of Fig.\;\ref{fig:SNR_all} in the $\Omega_{\rm m}$--$\sigma_8$ plane for UNIONS and Euclid respectively, this time finding the SNR of the difference of each cosmology with the fiducial combination $\Omega_{\rm m}=0.3$, $\sigma_8=0.85$. The SNR contours indicate how well a cosmology in this plane could be constrained  from the fiducial case. Also overlaid are recent results from the SRG$/$eROSITA all-sky survey \cite{eROSITA_abundance}, demonstrating the complementary degeneracy direction of constraints using the infall region.

 We can also consider each of the five (three) bins for Euclid (UNIONS) in Table \ref{tab:survey_bins} as independent tests, and sum their resulting SNRs in quadrature. For UNIONS, the improvement is marginal as the lower mass bins are noisier, and the cosmological dependence of their infall region is no longer orthogonal to $S_8$. On the other hand, for Euclid, the improvement in the SNR leads to a reduction in the contour area by a further factor of $\sim\!2$. The exact change in area is slightly uncertain, due to the interpolation between limited numbers of simulations (with parameters indicated by the open circles in Fig.\;\ref{fig:SNR}). Taken at face value, however, this measurement should reduce the area of the contour reported by eROSITA by a factor of more than $3$.

\section{Practical Challenges}
\label{sec:challenges}

\subsection{Use of a fiducial cosmology in the lensing calculations}
\label{sec:cosmo_assume}
Converting the observed shear profile $\gamma_t(\theta)$ into a projected $\Delta\Sigma(r)$ profile involves assuming a fiducial cosmology to convert angles and redshifts into distances. An incorrect choice of fiducial cosmology when converting angles to transverse distances causes an overall shift in scale; an incorrect choice of cosmology when calculating $\Sigma_{\rm crit}$ causes an overall shift in amplitude. Over the range of $\Omega_{\rm m}$ considered, the variation in the angular diameter distance with cosmology causes a $\pm 1.5\%$ shift in scale relative to Planck. Similarly, the variation of $\Sigma_{\rm crit}$ causes a $\pm 2.5\%$ change in the amplitude of the $\Delta\Sigma$ profile recovered, relative to Planck. Fortunately, these are competing effects. Assuming too high a value of $\Omega_{\rm m}$ when converting from $\gamma_t(\theta)$ to $\Delta\Sigma(r)$ would shift the profile to smaller $r$ and larger amplitude compared to the truth. Profiles from all simulations would be shifted in the same way, by different amounts depending on the value of $\Omega_{\rm m}$ relative to the fiducial case. Considering that the intrinsic variation in the profiles caused by cosmology is in a direction perpendicular to this shift, these two effects in combination largely cancel each-other out, leading to a $\leq 1\%$ impact on the resulting $\Delta\Sigma$ profile caused by a large mis-estimate of the true $\Omega_{\rm m}$. Thus, we do not include this correction in the main results.

\subsection{Measuring the correlation function \texorpdfstring{$\xi_{\rm mm}$}{ximm}}
\label{sec:xi}

To derive the bias profile corresponding to a given density profile, the matter-matter correlation function is required in Eq.\;(\ref{eq:bias}). This function is itself cosmology-dependent. It can be inferred from observational measurements of the galaxy correlation function, but then a model for the galaxy bias, connecting the distribution of galaxies to the distribution of matter, is needed. The galaxy bias is itself uncertain, and therefore including it introduces large uncertainties on the recovered cluster bias profile. Alternatively, instead of assuming a fixed $\xi_{\rm mm}(r)$, a cosmology dependent $\xi_{\rm mm}$ found using a halo model (or directly from simulations) could be included in the cosmological inference. This removes the need to estimate it from observation, but introduces extra freedom in the fitting procedure. Another consideration is that applying the same $\xi_{\rm mm}$ to the data and the simulation during the cosmological inference to generate a bias profile means we are essentially comparing the fitted density profile to a simulated density profile. This is similar to the method proposed in Section\;\ref{sec:fullprofile}, but includes the extra step of fitting the measured shear to get a density profile.

In the results presented in Section\;\ref{sec:depletion_results}, we naively assume the correlation function $\xi_{\rm mm}(r)$ for a Planck-like cosmology when deriving the bias profile, and  as a result, the true location of the bias minimum is not always recovered. Variations in $\xi_{\rm mm}$ are $\pm10\%$ in the infall region as $\sigma_8$ is varied by by $\pm0.05$. The variation is not constant with radius, so it changes the shape of the bias profile in the infall region, impacting the recovered depletion radius. Changing $\Omega_{\rm m}$ has a smaller but non-negligible impact. From this, we conclude that using the depletion radius (or minimum of the bias) as a cosmological probe is significantly hampered by uncertainty in the matter-matter correlation function, and thus probably not competitive with splashback or shear profile methods.

\subsection{Cluster mass calibration}

Mass calibration of galaxy clusters is a well known challenge. Accurate determinations of the mean mass and full mass distribution of the stacked halo sample are crucial in order to define equivalent samples of halos to compare to in simulations. Previous studies have shown there is a bias when using optically selected clusters---observed splashback radii are smaller than the values predicted by simulations \citep{sp_obs_10,sp_obs_16}. Euclid will take advantage of the weak lensing and spectroscopic data available for its mass calibration, as well as cross-correlating with data from other surveys \citep{Euclid_Clusters}. Furthermore, since we only consider the most massive clusters at low redshift, we can expect fairly complete samples. 

Nonetheless, Fig.\;14 of \cite{Euclid_mass_bias} shows that, in the redshift range of interest, average weak lensing cluster mass estimates may be biased low by $5$-$10\%$. We can investigate the impact of this bias by recomputing mean density and shear profiles for halos with $M\geq 0.9\times10^{14.3}$ and $M \geq 0.95\times10^{14.3}$, and comparing to the previous results obtained with $M \geq 10^{14.3}$, thus measuring how an uncorrected bias of $5-10\%$ would affect our final constraints. We find that the difference in the splashback radius in the stacked halo sample is typically $\ll 1\sigma$ in the case where all halo masses are biased low by $5\%$, where $\sigma$ is the uncertainty on the inferred value of $r_{\rm sp}$ from weak lensing profiles. When all halos are biased low by $10\%$, the difference is larger but still remains below the $1\sigma$ level. When comparing the amplitude of the stacked shear profile to simulations, errors are much smaller and the requirements on the mass calibration are more stringent. A downward mass bias of $10\%$ ($5\%$) for all halos reduces the lensing signal in Euclid clusters by $7\%$ ($3\%$), a difference that is significantly above the noise. While a mass bias of this amplitude is probably pessimistic, these results demonstrate the importance of accurate mass calibration. 

Furthermore, the mass uncertainty also has a significant impact. Halos with masses just below an imposed mass limit can be scattered into the sample, and those just above scattered out of the sample. The overall effect is a sample that contains less massive halos than expected. To test the impact, we generate 5 realisations of Euclid Bin 1 and UNIONS Bin 1 in Table\;1 with masses scattered from their true values assuming a mass uncertainty of $0.2\,$dex  \citep{Mass_scatter1,M_scatter0}. Results from these realisations are then compared to the original sample. As expected, the mean masses of the scattered realisations are each biased low by $\sim\!0.1\,$dex, and the measured splashback radii are biased low by $\sim\!1\sigma$. The shear profile amplitude is reduced by $10-15\%$ in the infall region. These results further highlight the importance of accurate mass calibration, especially when directly comparing the shear profiles to simulations. An improved method would include realistic mass scatter in the simulated halos when comparing to  observations, such that the main bias was accounted for. More sophisticated analysis methods, such as including abundance information by stacking a sample of the $N$ most massive clusters within a survey volume, or only selecting relaxed clusters \citep{relaxed1,relaxed2}, might also help reduce or avoid mass calibration problems. We will consider this possibility in future work.

\subsection{Mis-centering}
\label{sec:miscentre}

To assess the potential impact of stacking mis-centered halos on our study, we adopt the framework of \cite{offset}. The authors use a probability distribution for the size of the offset given by 
\begin{equation}
    P(R_{\rm off}) = \frac{R_{\rm off}}{\sigma_{\rm off}} \exp\left(-\frac{R_{\rm off}^2}{2\sigma_{\rm off}^2}\right) \, .
\end{equation}
Using a sample of galaxy groups detected in XMM-Newton \citep{XMM} and Chandra \citep{Chandra} observations, with $\Delta\Sigma$ profiles determined from COSMOS weak lensing measurements \citep{COSMOS_WL}, they find a best fit for the width of the offset distribution of $\sigma_{\rm off}=50\,$kpc. If we assume offsets are randomly drawn from this distribution for each individual system in a stacked halo sample, compute the modified $\Delta\Sigma(R)$ profile, and compare it to the case with no offsets, this will give an idea of the impact of mis-centering. 

We assume that every halo has an offset drawn from this distribution, a pessimistic choice corresponding to the worst-case scenario. For example, in the optically selected redMaPPer cluster sample, only $\sim\!25\%$ of clusters were mis-centered  \citep{redmapper}. Using $\sigma_{\rm off}=50\,$kpc, the difference in the amplitude of the shear profile can be up to $1$--$2\%$ in the infall region, and the impact on the recovered splashback radius can be a shift of $\sim\!0.05\,$Mpc$/h$. Both of these results are within the $1\sigma$ errors for Euclid shear measurements, and are sub-dominant for UNIONS measured profiles. Assuming $\sigma_{\rm off}$ can be estimated from observations, as in \cite{offset}, and only a fraction of clusters are strongly mis-centred, then we do not expect mis-centering to impact our conclusions. These results corroborate the findings of \cite{sp_obs_7}, where they evaluated the impact of mis-centering on the location of the 3D splashback radius, finding it to increase errors only slightly, causing shifts of the inferred location well within uncertainties.

\section{Conclusions}
 \label{sec:conclusions}

The amplitude and shape of the mean shear profile around galaxy clusters should depend on cosmology. In universes with low $\Omega_{\rm m}$ and high $\sigma_8$, clusters `form' earlier, assembling their mass into a single large progenitor at higher redshift relative to universes with high $\Omega_{\rm m}$ and low $\sigma_8$ \citep{Yuba}. Material accreted over the past few Gyr is accelerated by a deeper potential well, and reaches a larger splashback radius. Thus, the outer density profiles of clusters in low $\Omega_{\rm m}/$high $\sigma_8$ cosmologies will be more extended at the present day, and the shear signal in these regions will be larger. The dependence of this effect on $\Omega_{\rm m}$ and $\sigma_8$ is close to orthogonal to the $S_8$ degeneracy present in cosmic shear and cluster abundance studies (\hyperlink{citep.Roan}{H24}). Combining lensing measurements of the infall region around galaxy clusters with cluster number counts can therefore significantly improve constraints on these parameters, using only low-redshift cluster properties. 

Given realistic assumptions about the cluster samples and lensing data expected from Euclid, we find that measurements of the splashback radius alone may constrain deviations from fiducial Planck values of $\Omega_{\rm m}$ and $\sigma_8$ of $\pm0.05$ or greater. Measuring the `depletion radius', where the bias reaches a minimum, proves not to be a promising cosmological probe. This is due to the lack of a pronounced negative dip in the bias at this radius on cluster scales, and also the need to determine the matter-matter correlation function in order to calculate the bias. A similar test could in principle be applied to lower-mass groups or even galaxy halos. In particular, the depletion radius is a more prominent feature in low-mass dark matter halos, and thus easier to measure. The density profiles of these lower-mass systems are more likely to be impacted by baryonic feedback effects, however, given their lower energy scale, as demonstrated by small-scale lensing or clustering studies \cite[e.g.][]{Huang19}. Using the minimum bias value, instead of the radius at which the bias reaches a minimum, might improve the prospects of the depletion radius as a cosmological probe for galaxy clusters, although it requires a good knowledge of the matter density of and the matter-matter correlation function.

Furthermore, we find that variation of infall region features with cosmology in these halos is no longer orthogonal to $S_8$ as with galaxy clusters, but has a similar degeneracy direction to $S_8$. This limits their utility in improving abundance or lensing-based constraints. Based on these arguments and our exploration of the SNR of differences in shear profiles, we can conclude the best halos to use for a cosmological analysis are high-mass clusters at low redshift.

Fitting analytic forms to observed shear profiles to infer $r_{\rm sp}$ or $r_{\rm cd}$ is also a slightly noisy process, reducing the overall SNR of these tests. By applying a simple angular filter to the mean shear profiles in the infall region for our simulated clusters, and studying the intrinsic variation of the mean shear with cosmology, we demonstrated that direct measurements of the profile in weak lensing surveys such as UNIONS or Euclid can better constrain $\Omega_{\rm m}$ and $\sigma_8$, reaching a final precision  comparable to that of cluster abundance studies, but with a different degeneracy direction in the $\Omega_{\rm m}$--$\sigma_8$ plane. These constraints could be further improved by using an optimal weighted filter to extract the largest SNR from the infall region  \citep[e.g.][]{cluster_lensing_SNR}.

The real power comes from combining constraints based on measurements of the density profile with traditional results based on cluster abundance. We estimate that combining the two could reduce the area of the contours reported by the SRG$/$eROSITA All-Sky Survey by a factor of $1.2$, using one mass and redshift bin from the UNIONS survey, or a factor of $3$, using five redshift bins from the Euclid survey. 
Overall, using the full range of cluster mass and redshift available, lensing measurements of the infall zone should provide a competitive, independent cosmological test that is  highly complementary to other low-redshift tests of cosmology, and requires only the data already collected for abundance and cosmic shear studies. In a follow-up work we will attempt the first such analysis using clusters in the UNIONS footprint.

\section*{Acknowledgements}

We thank the members of the UNIONS collaboration and D. Rana for helpful feedback in the preparation of this manuscript. C. T. M. is funded by a Leverhulme Trust Study Abroad Scholarship. J. E. T. acknowledges support from the Natural Sciences and Engineering Research Council of Canada (NSERC), through a Discovery Grant. This research was enabled in part by support provided by Compute Ontario  (www.computeontario.ca) and the Digital Research Alliance of Canada (alliancecan.ca). The python packages \textsc{numpy}, \textsc{scipy}, \textsc{matplotlib}, \textsc{lmfit}, \textsc{colossus} and \textsc{cluster-lensing-cov} have been used in this work. For the purpose of open access, the authors have applied a Creative Commons Attribution (CC BY) licence to any Author Accepted Manuscript version arising from this submission.

%%%%%%%%%%%%%%%%%%%%%%%%%%%%%%%%%%%%%%%%%%%%%%%%%%
\section*{Data Availability}

The simulations used in this article will be shared on reasonable request to the corresponding author.


\bibliographystyle{mnras}
\bibliography{infall} 

%%%%%%%%%%%%%%%%%%%%%%%%%%%%%%%%%%%%%%%%%%%%%%%%%%

%%%%%%%%%%%%%%%%% APPENDICES %%%%%%%%%%%%%%%%%%%%%

\appendix

\section{Variation of the characteristic radii with redshift}
\label{app:z_var}

\begin{figure*}
    \centering
   \includegraphics[width=\linewidth]{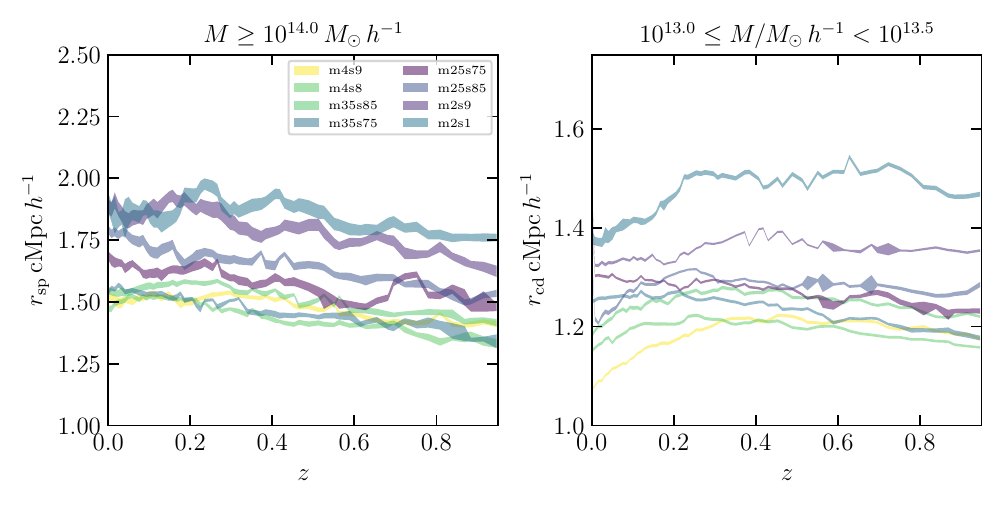}
    \caption{The mean splashback (left) and depletion (right) radii in comoving units, for samples in the mass ranges indicated, as a function of redshift and cosmology.}
    \label{fig:r_with_z}
\end{figure*}

Although it is tangential to our main argument, it is also interesting to study how the cosmology dependence of the characteristic radii varies with redshift. Fig.\;\ref{fig:r_with_z} shows the redshift dependence of the splashback and depletion radii for a sub-sample of cosmologies. We see that in comoving units, there is relatively little variation ($\leq$ 20\%) between $z =0$ and $z=1$. This agrees with other works, \citep[e.g.][]{Diemer2_cosmo,Splashback_sims}, where the authors find any variation of the splashback radius with redshift is caused by the variation of $\Omega_{\rm m}(z)$, meaning that little variation is expected in comoving units.

\section{Dependence of errors on cosmology}
\label{app:err_cosmo}

\begin{figure}
    \centering
    \includegraphics[width=\linewidth]{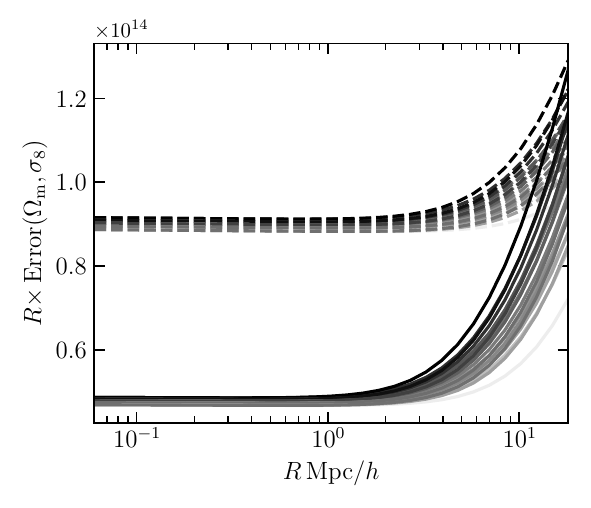}
    \caption{Error on $\Delta\Sigma(R)$ as a function of combinations of cosmology. Dashed lines correspond to clusters observed in UNIONS, and solid lines to clusters observed in Euclid. Darker line colours correspond to larger values of  $S_8$.}
    \label{fig:err_cosmo}
\end{figure}

Due to the presence of the matter power spectrum in the covariance calculation, the predicted errors are in principle cosmology dependent. 
In our analysis, we make the approximation that the errors are equal to those of a fiducial Planck-like cosmology with $\Omega_{\rm m}=0.316$ and $\sigma_8=0.811$ \citep{Planck}. Fig.\;\ref{fig:err_cosmo} shows how the actual uncertainties depend on cosmology, for Euclid Bin 1 (solid lines) and UNIONS Bin 2 (dashed lines) defined in Table \ref{tab:survey_bins}. The uncertainty is plotted as a function of $\Omega_{\rm m}$ and $\sigma_8$, with darker line colours corresponding to larger values of $S_8$. The variation is seen to be relatively small in the infall region, so we have neglected this correction in our calculations. In practice there are ways of dealing with this cosmology dependence in a likelihood analysis \citep[e.g.][]{Like_cosmo}, thereby slightly improving the accuracy of the SNR calculations. 

%%%%%%%%%%%%%%%%%%%%%%%%%%%%%%%%%%%%%%%%%%%%%%%%%%

% Don't change these lines
\bsp	% typesetting comment
\label{lastpage}
\end{document}